\documentclass[letterpaper, 10pt, conference]{ieeeconf}      % Use this line for a4
\IEEEoverridecommandlockouts                              % This command is only
\newcommand\bfhat[1]{\mathbf{\hat{#1}}}
\newcommand{\argmin}{\operatornamewithlimits{argmin}}
\newcommand{\qed}{\hfill  \rule{2mm}{3mm}}

\usepackage{amsmath,amssymb,graphics,epsfig,latexsym,color}
\usepackage{times}
\usepackage{subfigure}
\usepackage{psfrag}
\newcounter{mytempeqncnt}

\newtheorem{lemma}{Lemma}
\newtheorem{theorem}[lemma]{Theorem}

\newcommand{\junk}[1]{}

\title{\LARGE \bf
On successive refinement of diversity  for fading ISI channels
}

\author{S. Dusad and S. N. Diggavi% <-this % stops a space 
  \thanks{EPFL, Lausanne, Switzerland, S. Dusad was supported in part by SNSF
  Grant \# 200021-105640/1.  S. N. Diggavi is part of the SNSF
  supported NCCR-MICS center on wireless sensor networks.  Email:
  \{suhas.diggavi,sanket.dusad\}@epfl.ch.  } }

\begin{document}

\maketitle
\thispagestyle{empty}
\pagestyle{empty}

%%%%%%%%%%%%%%%%%%%%%%%%%%%%%%%%%%%%%%%%%%%%%%%%%%%%%%%%%%%%%%%%%%%%%%%%%%%%%%%%
\begin{abstract}

Rate and diversity impose a fundamental trade-off in
communications. This tradeoff was investigated for Inter-symbol
Interference (ISI) channels in \cite{dtse}.  A different point of view
was explored in \cite{dimacs} where high-rate codes were designed so
that they have a high-diversity code embedded within them. Such
diversity embedded codes were investigated for flat fading channels
and in this paper we explore its application to ISI channels. In
particular, we investigate the rate tuples achievable for diversity
embedded codes for scalar ISI channels through particular coding
strategies. The main result of this paper is that the diversity
multiplexing tradeoff for fading ISI channels is indeed successively
refinable. This implies that for fading single input single output
(SISO) ISI channels one can embed a high diversity code within a high
rate code without any performance loss (asymptotically). This is
related to a deterministic structural observation about the asymptotic
behavior of frequency response of channel with respect to fading
strength of time domain taps.

\end{abstract}

%%%%%%%%%%%%%%%%%%%%%%%%%%%%%%%%%%%%%%%%%%%%%%%%%%%%%%%%%%%%%%%%%%%%%%%%%%%%%%%%

\section{Introduction}
\label{sec:intro}
There exists a fundamental tradeoff between diversity (error
probability) and multiplexing (rate). This tradeoff was characterized
in the high SNR regime for flat fading channels with multiple transmit
and multiple receive antennas (MIMO) \cite{ZhengTse02b}. This
characterization was done in terms of multiplexing rate which
captured the rate-growth (with $SNR$) and diversity order which
represented reliability (at high $SNR$). This diversity multiplexing
(D-M) tradeoff has been extended to several cases including fading ISI
channels \cite{dtse}, \cite{medles}. The presence of ISI gives
significant improvement of the diversity order. In fact, for the SISO
case the improvement was equivalent to having multiple receive
antennas equal to the number of ISI taps \cite{dtse}.

A different perspective for opportunistic communication was presented
in \cite{dimacs}, \cite{suhas}. A strategy that combined high rate
communications with high reliability (diversity) was
investigated. Clearly, the overall code will still be governed by the
rate-reliability tradeoff, but the idea was to ensure the high
reliability (diversity) of at least part of the total
information. These are called diversity-embedded codes
\cite{dimacs}, \cite{suhas}. In \cite{suhas} it was shown that when we
have one degree of freedom (one transmit many receive or one receive
many transmit antennas) the D-M tradeoff was successively
refinable. That is, the high priority scheme (with higher diversity
order) can attain the optimal diversity-multiplexing  (D-M) performance as if
the low priority stream was absent. However, the low priority scheme
(with lower diversity order) attains the same D-M performance as that
of the aggregate rate of the two streams. When there is more than one
degree of freedom (for example, parallel fading channels) such a
successive refinement property does not hold \cite{itw06}.

In this paper we investigate the diversity embedded codes for an ISI
channel with single transmit and receive antenna. Since the Fourier
basis is the eigenbasis for linear time invariant channels we can
decompose the transmission into a set of parallel channels. Since it
is known that the D-M tradeoff for parallel fading channels is not
successively refinable \cite{itw06}, it is tempting to expect the same
for fading ISI channels. However, the main result of this paper is
that for SISO fading ISI channels the D-M tradeoff is indeed
successively refinable. The correlations of the fading across the
parallel channels seem to cause the difference in the behavior.  The
structural observations in lemma \ref{magnitude} give insight into
these correlations. This will be made more precise in the paper.

The paper is organized as follows. In Section 2 we formulate the
problem statement and present the notation. Section 3 gives a
variation of the proof in \cite{dtse} of the D-M tradeoff for ISI
channels which makes a connection to the diversity embedded codes. We
explore the role of correlation in successive refinement through a
specific example. Section 4 presents the statement and the proof for
the successive refinability of the D-M tradeoff for ISI channels. We
conclude the paper with a brief discussion followed by the details of
the proofs in the appendix.

\section{Problem Statement}

Consider communication over a quasi static fading channel with
Inter-symbol Interference (ISI)
\begin{equation}
\label{eqn:sysmodel}
y[n] = h_0 x[n] + h_1 x[n-1] +\ldots +h_{\nu}x[n-\nu] + z[n]
\end{equation}
The $\nu+1$ i.i.d. fading coefficients are $h_i \sim
\mathcal{C}\mathcal{N}(0,1)$ and fixed for the duration of the block
length ($N+\nu$). The additive noise $z[n]$ is i.i.d. circularly
symmetric Gaussian with unit variance. As is standard in these
problems, we assume perfect channel knowledge only at the receiver.

The coding scheme is limited to one quasi-static transmission block of
size $N+\nu$. Consider a sequence of coding schemes with transmission
rate as a function of $SNR$ given by $R(SNR)$ and an average error
probability of decoding ${P}_e(SNR)$. Analogous to \cite{ZhengTse02b} we
define the multiplexing rate $r$ and the diversity order $d$ as
follows,
\begin{equation}
\label{eq:DivTuple} d =\lim_{SNR\rightarrow\infty}
-\frac{\log {P}_e(SNR)}{\log(SNR)} ,\,\, r
=\lim_{SNR\rightarrow\infty}
\frac{R(SNR)}{\log(SNR)}.
\end{equation}

With these definitions, the D-M tradeoff for ISI channels was
established in \cite{dtse}. 
\begin{theorem}{\cite{dtse}}
\label{theorem:dtse}
The diversity multiplexing tradeoff for the system model in
(\ref{eqn:sysmodel}) is bounded by,
\begin{align}
(\nu+1)\left(1-\frac{N+\nu}{N}r \right) & \leq d_{isi}(r) \leq  (\nu+1)\left(1-r\right) \label{eqn:isitradeoff}
\end{align}
\end{theorem}
In this paper we explore the performance of diversity embedded codes
over ISI channels. For clarity we focus on two streams but the
procedure can be generalized to more than two levels.

Let $\mathcal{H}$ denote the message set from the first information
stream and $\mathcal{L}$ denote that from the second information
stream. The rates for the two message sets as a function of $SNR$ are,
respectively, $R_H(SNR)$ and $R_L(SNR)$. The decoder jointly decodes
the two message sets and we can define two error probabilities,
$P_e^H(SNR)$ and $P_e^L(SNR)$, which denote the average error
probabilities for message sets $\mathcal{H}$ and $\mathcal{L}$
respectively. We want to characterize the tuple $(r_H, d_H, r_L, d_L)$
of rates and diversities for the ISI channel that are achievable,
where analogous to (\ref{eq:DivTuple}),
\begin{align*}
\label{eq:EmbDivTuple} 
d_H =\lim_{SNR\rightarrow\infty}
-\frac{\log {P}_e^H(SNR)}{\log(SNR)},& \,\,  r_H
=\lim_{SNR\rightarrow\infty}
\frac{R_H(SNR)}{\log(SNR)} \\
d_L =\lim_{SNR\rightarrow\infty}
-\frac{\log {P}_e^L(SNR)}{\log(SNR)},& \,\,  r_L
=\lim_{SNR\rightarrow\infty}
\frac{R_L(SNR)}{\log(SNR)} 
\end{align*}
Also, we assume that $d_H\geq d_L$.  Note that for the joint codebook
$\{ \mathcal{H}, \mathcal{L}\}$ the total multiplexing rate is
$r_H+r_L$ and the diversity $d=min(d_H,d_L)=d_L$. We use the special
symbol $\doteq$ to denote exponential equality {\em i.e.}, we write $f(SNR)\doteq SNR^b$ to denote
\begin{align*}
\lim_{SNR\rightarrow\infty}\frac{\log f (SNR)}{\log(SNR)} & =  b 
\end{align*}
and $\stackrel{\cdot}{\leq}$ and $\stackrel{\cdot}{\geq}$ are defined similarly.

From an information-theoretic point of view \cite{suhas} focused on
the case when there is one degree of freedom ({\em i.e.,}
$\min(M_t,M_r)=1$). In that case if we consider $d_H \geq d_L$ without
loss of generality, the following result was established in
\cite{suhas}.

\begin{theorem}
\label{thm:SucRef}
When $\min(M_t,M_r)=1$, then the diversity-multiplexing trade-off
curve is successively refinable, i.e., for any multiplexing rates
$r_H$ and $r_L$ such that $r_H+r_L \leq 1$, the diversity orders
$d_H\geq d_L$,
\begin{align}
%\label{eq:SucRef}
d_H & = d^{opt}(r_H), \,\,\,\\
d_L & = d^{opt}(r_H+r_L)
\end{align}
are achievable, where $d^{opt}(r)$ is the optimal diversity order
given in \cite{ZhengTse02b}.
\end{theorem}
\hfill $\blacksquare$

Since the overall code has to still be governed by the rate-diversity
trade-off given in \cite{ZhengTse02b}, it is clear that the trivial
outer bound to the problem is that $d_H\leq d^{opt}(r_H)$ and $d_L\leq
d^{opt}(r_H+r_L)$.  Hence Theorem \ref{thm:SucRef} shows that the best
possible performance can be achieved.  This means that for
$\min(M_t,M_r)=1$, we can design ideal {\em opportunistic} codes. This
analysis was done for flat fading channels and we will show a similar
theorem for ISI fading channels.

\section{ISI Tradeoff}

In this section we give an alternative interpretation of the D-M
tradeoff for ISI channels for the particular case of two taps {\em
i.e.}, $\nu=1$. This exercise will help us to see the difference
between the fading ISI channel and the i.i.d. parallel fading channel models.

Rewriting the equation (\ref{eqn:sysmodel}) for the case of
two taps, we have,
\begin{equation}
\label{eqn:sysmodel2tap}
y[n] = h_0 x[n] + h_1 x[n-1] + z[n]
\end{equation}
Assume a scheme in which one data symbol is sent in every $(\nu+1)$
transmissions from a QAM constellation of size $SNR^r$. With this
strategy there is no interference between successive transmitted
symbols and the receiver performs matched filtering to recover the
symbol from the $\nu+1$ copies of the received signal. This gives us
the matched filter upper bound to the diversity or the lower bound to
the error probability,
\begin{align*}
d_{isi}(r) & \leq   2(1-r)
\end{align*}
where $r$ is the multiplexing rate and $d$ is the diversity order.

For the lower bound to the diversity consider a transmission strategy
in which we assume that after transmission over a block length $N$, in
the last $\nu=1$ instants zero symbol is transmitted in order to
avoid interblock interference. Therefore, the received vector over the
block of length $N+\nu$ can be written as,
\begin{align}
\label{circulant_model}
\left[\begin{array}{c} y[0]\\  y[1] \\ \vdots \\ y[N] \end{array} \right]
 & =  \underbrace{\left[
\begin{array}{cccccc}
h_0    & 0      & \ldots & \ldots & h_1 \\
h_1    & h_0    & \ldots & \ldots & 0   \\
\vdots & \vdots & \ddots & 0      & 0   \\
0      & \ldots & h_1    & h_0    & 0   \\
0      & \ldots & 0      & h_1    & h_0  
\end{array}
\right]}_{\mathbf{H}}
\left[ \begin{array}{c}
x[0] \\ x[1] \\ \vdots \\ x[N-1] \\ 0 
\end{array}
\right]
+\mathbf{z}
\end{align}
where $\mathbf{z}=\left[ \begin{array}{cccc}z[0] & z[1] & \ldots &
z[N] \end{array} \right]^T$. Note now that the channel matrix
$\mathbf{H}$ is a circulant matrix. Proceeding as in \cite{dtse}, look
at the circulant matrix $\mathbf{H=Q}\mathbf{\Lambda}\mathbf{ Q^*}$ in
the frequency domain where ${\bf Q}$ and ${\bf Q^*}$ are truncated DFT
matrices and $\mathbf{\Lambda}$ is a diagonal matrix with the diagonal
elements given by $\Lambda_l=h_0+h_1 e^{-\frac{l 2 \pi j}{N+1}}$ for
$l=0,\ldots,N$.  If $(N+1)$ is divisible by two,we can view this as
$\frac{(N+1)}{2}$ sets of $2$ parallel independent
channels. Communicating over each such set of $2$ parallel channels at
a rate $2r$, we get the effective multiplexing rate,
$\tilde{r}=r\frac{N}{N+1}$. The diversity for this multiplexing rate
is $2-2r=2(1-\frac{N+1}{N}\tilde{r})$ as given in \cite{ZhengTse02b}.

{\em Note that in the above argument we do not utilize the fact that
correlations exist across these sets of independent channels
anywhere. In this case correlations did not matter since it achieves\footnote{asymptotically in $N$.}
the matched filter upper bound.}

In order to illustrate the impact of correlation between the
frequency domain coefficients, we will specifically consider the case
for $N=3$ and $\nu=1$. Using (\ref{circulant_model}) and the Fourier
decomposition we see that we have four parallel channels,
\begin{align*}
\tilde{y}_l & =  \Lambda_l \tilde{x}_l + \tilde{z}_l \qquad l=0,\ldots,3
\end{align*}
In the spirit of the above parallel channel argument we can view these
as two sets of parallel channels $\{\tilde{y}_0 , \tilde{y}_2\}$ and
$\{\tilde{y}_1 , \tilde{y}_3\}$ each consisting of two
sub-channels. Given this view since the D-M tradeoff for parallel
channels are not successively refinable we would expect that such a
characterization also hold for the ISI D-M tradeoff. However, since $
\Lambda_0 = h_0 + h_1$, $\Lambda_2 = h_0 - h_1$, $\Lambda_1 = h_0 -j
h_1$, $\Lambda_3 = h_0 +j h_1$, we see that the fading across the two
sets of parallel channels are correlated. In particular, if
$|h_0|^2\stackrel{\cdot}{>}SNR^{-(1-r)}$ then asymptotically
$|\Lambda_l|^2 \stackrel{\cdot}{\leq} SNR^{-(1-r)}$ for at most one
$l\in\{0,\ldots,3\}$. Therefore, it is possible to code across these
sets of parallel channels to get better performance instead of
treating them independently.

This example gives the intuition to use the following method to prove
the diversity multiplexing tradeoff for the ISI channel.  Define a set
$\mathcal{A}$ of events such that
\begin{align}
\mathcal{A}& =  \left\{ \mathbf{h} :  |h_0|^2 \stackrel{\cdot}{\leq}\frac{1}{SNR^{1-r}} \ {\rm and} \ |h_1|^2 \stackrel{\cdot}{\leq}\frac{1}{SNR^{1-r}}  \right\} 
\end{align}
For high $SNR$ it follows that the probability of the set
$\mathcal{A}$ occurring is $P(\mathcal{A})=SNR^{-2(1-r)}$ and
$P(\mathcal{A}^c)=1-SNR^{-2(1-r)}$.  At each time instant
independently transmit one symbol from a constellation with
$d_{min}^2\stackrel{\cdot}{\geq}SNR^{(1-r)}$ for $N$ time instants and
pad it with $\nu$ zero symbols. For detection, given that $\mathbf{h}
\in \mathcal{A}^c $, we proceed as in the proof of lemma 3 with the
detection of the $N+\nu$ length transmitted sequence in the frequency
domain. Clearly, the error probability of the scheme with this
decoder, denoted by $ P_e^{D}(SNR)$, is an upper bound to the error
probability {\em i.e.}, $P_e(SNR) \leq P_e^{D}(SNR)$. Therefore, we
can write,
\begin{align*}
\lefteqn{P_e(SNR) =   P(\mathcal{A}) P_{e}(SNR \mid  \mathbf{h} \in \mathcal{A} ) +} & \\
& \hspace{0.2in} P(\mathcal{A}^c) P_{e}(SNR\mid \mathbf{h} \in \mathcal{A}^c ) \\
& \stackrel{\cdot}{\leq}   P(\mathcal{A}) + \left( 1-SNR^{-2(1-r)} \right)  P_{e}(SNR\mid \mathbf{h} \in \mathcal{A}^c ) \label{errorprob} \\
& \stackrel{\cdot}{\leq}   SNR^{-2(1-r)} + \left( 1-SNR^{-2(1-r)} \right)   P_{e}^{D}(SNR\mid \mathbf{h} \in \mathcal{A}^c ) \\
& \doteq   SNR^{-2(1-r)}.
\end{align*}
The last equality is true due to lemma \ref{lemma1} (given in Section
4) which states that if $\mathbf{h}\in\mathcal{A}^c$ the probability
of error decays exponentially in $\mathrm{SNR}$. This lemma in turn is
based on a structural observation made in lemma \ref{magnitude} (also
see Section 3) that at most $\nu$ coefficients in frequency domain
will be smaller than $\min_{l} |h_l|^2$ and here given that
$\mathbf{h}\in\mathcal{A}^c$ at most $\nu$ coefficients will be
smaller than $SNR^{-(1-r)}$. This method of analysis turns out to be
more useful for us and takes into account the fact that the sets of
parallel channels are correlated.

%For successive refinability we use superposition codes with the high
%priority scheme having power $P_H$ and the low priority scheme having
%power $P_L$ such that $SNR_H\stackrel{\cdot}{=}SNR$ and
%$SNR_L\stackrel{\cdot}{=} SNR^{1-\beta}$.

This example was specifically for $N=3$ and 2 taps. A similar analysis
carries over for the case of general $N$ and $\nu$.  As summarized in
lemma 2 in Section 4, for a finite $N$ with $(\nu +1)$ taps either all
the taps will be of order less than $SNR^{-(1- r)}$ or at most $\nu$
taps will be of order less than $SNR^{-(1- r)}$.

\section{Successive Refinement of the ISI D-M tradeoff}

In this section we will formally prove the successive refinement of
the D-M tradeoff for ISI channels. The intuition of the effect of
fading in the frequency domain is captured by the following result
which is proved in the appendix.

\begin{lemma}
\label{magnitude}
For a $(\nu+1)$ tap
ISI channel we have the taps in the frequency domain are given by,
\begin{align*}
\Lambda_{k} & =   \sum_{m=0}^{\nu} h_m e^{-\frac{2 \pi j}{(N+\nu)}k m} \qquad k =\{0,\ldots,(N+\nu-1)\}
\end{align*}
Define the sets $\mathcal{F}$, $\mathcal{G}$  and $\mathcal{A}$ as,
\begin{align} 
\mathcal{F} & =  \{i : |\Lambda_i|^2\stackrel{\cdot}{\leq}
SNR^{-(1-r)}\},\\
\mathcal{G} & =\{i : |\Lambda_i|^2 \doteq
\max_{l\in\{ 0,1,\ldots,\nu\} }|h_l|^2\} , \nonumber \\ 
\mathcal{A} & =  \left\{ \mathbf{h} : |h_m|^2
\stackrel{\cdot}{\leq}\frac{1}{SNR^{1-r}} \quad \forall m \in
\{0,1,\ldots,\nu\} \right\}
\end{align}
With these definitions we have:
\renewcommand\theenumi{(\alph{enumi})}
\begin{enumerate} 
\item Given that $\mathbf{h} \in \mathcal{A}^c$, $|\mathcal{F}|\leq
\nu$, i.e., at most $\nu$ taps in the frequency domain are
(asymptotically) of magnitude less than $SNR^{-(1-r)}$.
\item $|\mathcal{G}^c|\leq \nu $ i.e., at least $N$ taps of the
$N+\nu$ taps in the frequency domain are (asymptotically) of magnitude
$\max(|h_0|^2,|h_1|^2,\ldots,|h_\nu|^2)$,
\begin{align*}
|\{k:|\Lambda_{k}|^2  \stackrel{\cdot}{<}  \max(|h_0|^2,|h_1|^2,\ldots,|h_\nu|^2) \}| & \leq \nu.
\end{align*}
\end{enumerate}
\qed
\end{lemma}
Note that $(b)$ along with $\mathbf{h}\in\mathcal{A}^c$ implies $(a)$
and therefore is the stronger claim. Here is an intuition of why such
a result will hold. Consider the polynomial
\begin{align}
\Lambda(z)  & =   \sum_{m=0}^{\nu} h_m z^{m},
\nonumber
\end{align}
which evaluates to the Fourier transform for $z=e^{-\frac{2 \pi
j}{(N+\nu)}k}$. Hence, if we evaluate the polynomial at $z=e^{-\frac{2
\pi j}{(N+\nu)}k}$, for $k =\{0,\ldots,(N+\nu-1)\}$, at most $\nu$
values can be zero and at least $N$ values are bounded away from
zero. Therefore, if $SNR$ is large enough, it is clear that at least
$N$ values would be ``larger'' than $SNR^{-(1-r)}$. The details of the
proof are in the appendix.

Now consider transmission using uncoded QAM such that the minimum
distance between any two points in the constellation $d_{min}$ is such
that $d_{min}^2\stackrel{\cdot}{\geq} SNR^{(1-r)}$.  Defining
$\mathcal{F}=\{i : |\Lambda_i|^2\stackrel{\cdot}{\leq} SNR^{-(1-r)}\}$ we have from the
lemma above that $|\mathcal{F}|\leq \nu$. Ignore these $\nu$ channels
and examine the remaining $N$ channels in $\mathcal{F}^c$. We can show
that the distance between codewords in these channels is still
asymptotically larger than $SNR^{(1-r)}$. Since the pairwise error
probability is a $Q$ function, we can show that the error probability
decays exponentially in SNR. This is summarized in the following
lemma, the proof of which is in the appendix.
\begin{lemma}
\label{lemma1}
Assume that the minimum distance $d_{min}$ between any two points in
the constellation ($\mathcal{X}$) from which the signal is transmitted
is $d_{min}^2\stackrel{\cdot}{\geq} SNR^{(1-r)}$. Assume uncoded
transmission such that at each time instant one symbol is
independently transmitted from the constellation for $N$ time instants
followed by a padding with $\nu$ zero symbols. For a finite period of
communication (finite $N$) given that $\mathbf{h} \in \mathcal{A}^c$
(see (7)), the error probability $P_e$ decays exponentially in $SNR$.
\qed
\end{lemma}

The part $(a)$ of lemma \ref{magnitude} and lemma \ref{lemma1} can be
combined together to give an alternative proof of the diversity
multiplexing tradeoff of the ISI channel. But to prove the successive
refinement of the D-M tradeoff of the ISI channel we need the stronger
result in the part $(b)$ of lemma \ref{magnitude}.

We will prove a lemma analogous to lemma \ref{lemma1} for the case of
superposition coding. This will be useful in our proof to show the
successive refinement of the D-M tradeoff for ISI channels. Since we
are padding every $N$ symbols with $\nu$ zeros, to communicate at an
effective rate of $r$ using uncoded QAM transmission, we need to send
symbols from a QAM constellation of size $SNR^{\tilde{r}}$ where
$\tilde{r}=\frac{r(N+\nu)}{N}$. Let $\mathcal{X}_H$ be QAM
constellation instant of size $SNR^{\tilde{r}_H}$ and power constraint
$SNR$. Similarly let $\mathcal{X}_L$ be a QAM constellation of size
$SNR^{\tilde{r}_L}$ and power constraint $SNR^{1-\beta}$, where
$\beta>\tilde{r}_H$. As before, define
\begin{align}
\label{A_high}
\mathcal{A}_H & =   \left\{ \mathbf{h} :  |h_m|^2 \stackrel{\cdot}{\leq}\frac{1}{SNR^{1-\tilde{r}_H}} \quad \forall m \in \{0,1,\ldots,\nu\}  \right\} 
\end{align}

\begin{lemma}
\label{lemma3}
Using the $\mathcal{X}_H$ and $\mathcal{X}_L$ for signaling, assume
uncoded superposition transmission such that at each time instant
symbols are independently chosen and superposed from each
constellation ($\mathcal{X}_H$, $\mathcal{X}_L$) for $N$ time instants
followed by a padding with $\nu$ zero symbols. For a finite period of
communication (finite $N$) given that $\mathbf{h} \in \mathcal{A}_H^c$
(see (\ref{A_high})), the error probability of detecting the set of
symbols sent from the higher constellation ($\mathcal{X}_H$) denoted
by $P_e^{H}(SNR)$ decays exponentially in $SNR$.  \qed
\end{lemma}
In this lemma we critically use the fact that {\em all} except at most
$\nu$ taps in the frequency domain, are asymptotically of {\em equal}
magnitude ($\max_{l\in\{ 0,1,\ldots,\nu\} }|h_l|^2$).

Using these lemmas we will prove the following theorem on the
successive refinement.
\begin{theorem}
\label{theorem_siso}
Consider a $\nu$ tap point to point SISO ISI channel. The diversity
multiplexing tradeoff for this channel is successively refinable, {\em
i.e.}, for any multiplexing gains $r_H$ and $r_L$ such that $r_H+r_L
\leq \frac{N}{N+\nu}$ the achievable diversity orders given by
$d_{H}(r_H)$ and $d_{L}(r_L)$ are bounded as,
\begin{align}
(\nu+1)\left(1-\frac{N+\nu}{N}r_H\right)  & \leq  d_{H}(r_H) \nonumber \\
& \hspace{-0.4in} \leq  (\nu+1)\left(1-{r_H}\right) \label{higherlayer}, \\
(\nu+1)\left(1-\frac{N+\nu}{N}(r_{H}+r_L)\right) & \leq   d_{L}(r_{L}) \nonumber \\
&\hspace{-0.4in} \leq  (\nu+1)\left(1-(r_H+r_L)\right)  \label{higherlower}
\end{align}
where $N$ is finite and does not grow with SNR. 
\qed
\end{theorem}
\begin{proof}
To show the successive refinement we use superposition coding and
assume two streams with uncoded QAM codebooks for each stream, as in
\cite{suhas}. Assume that given a total power constraint $P$ we
allocate powers $P_{H}$ and $P_{L}$ to the high and low priority
streams respectively. We design the power allocation such that at high
signal to noise ratio, we have $SNR_{H}\doteq SNR$ and $SNR_{L} \doteq
SNR^{1-\beta}$ for $\beta \in [0,1]$. Let $\mathcal{X}_H$ be QAM
constellation instant of size $SNR^{\tilde{r}_H}$ with minimum
distance $(d_{min}^H)^2=SNR^{1-\tilde{r}_H}$. Similarly let
$\mathcal{X}_L$ be a QAM constellation of size $SNR^{\tilde{r}_L}$
with minimum distance $(d_{min}^L)^2=SNR^{1-\beta-\tilde{r}_L}$,
where $\beta>\tilde{r}_H$. The symbol transmitted at the $k^{th}$
instant is the superposition of a symbol from $\mathcal{X}_H$,
$\mathcal{X}_L$ given by,
\begin{align*}
x[k] & =  x_H[k]+x_L[k] \qquad {\rm where} \, \, x_H[k] \in \mathcal{X}_H,  \,\, x_L[k] \in \mathcal{X}_L
\end{align*}
It can be shown \cite{suhas} that even with the above superposition
coding, if $\beta>\tilde{r}_H$ the order of magnitude of the effective
minimum distance between two points in the constellation
$\mathcal{X}_H$ is preserved.

The upper bound in both (\ref{higherlayer}) and (\ref{higherlower}) is
trivial and follows from the matched filter bound. We will investigate
the lower bound in (\ref{higherlayer}). At each time instant superpose
symbols from the higher and lower layers for $N$ time instants and pad
them with $\nu$ zero symbols at the end.  We consider this particular
transmission scheme and for detection, given that $\mathbf{h} \in
\mathcal{A}^c_H $, (where $\mathcal{A}_H$ is as defined in equation
(\ref{A_high})), we proceed as in lemma 3. Therefore, we can write,
\begin{align}
\lefteqn{P_{e}^H(SNR) =  P(\mathcal{A}_H) P_{e}(SNR \mid \mathbf{h} \in \mathcal{A}_H ) + }& \nonumber \\
& \hspace{0.2in} P(\mathcal{A}^c_H) P_{e}(SNR\mid \mathbf{h} \in \mathcal{A}^c_H )  \\
& \stackrel{\cdot}{\leq}   P(\mathcal{A}_H) + \left( 1-SNR^{-(\nu+1)(1-\tilde{r}_H)} \right)  P_{e}(SNR\mid \mathbf{h} \in \mathcal{A}^c_H )  \nonumber \\
& \stackrel{\cdot}{\leq}   SNR^{-(\nu+1)(1-\tilde{r}_H)} +\nonumber\\
& \hspace{0.2in} \left( 1-SNR^{-(\nu+1)(1-\tilde{r}_H)} \right)   P_{e}^{D}(SNR\mid \mathbf{h} \in \mathcal{A}^c_H ) \label{errorprob2}
\end{align}
for communication at an effective rate of $r_H=\frac{N}{N+\nu}\tilde{r}_H$.

\begin{figure*}[!t]
\normalsize
\setcounter{mytempeqncnt}{\value{equation}}
\setcounter{equation}{21}
\begin{align}
\label{circulant_model_simo}
\left[\begin{array}{c} \mathbf{y}[0]\\  \mathbf{y} [1] \\ \vdots \\ \mathbf{y}[N]  \\  \ldots \\ \mathbf{y}[N+\nu] \end{array} \right]
 & = 
\underbrace{
\left[
\begin{array}{cccccccc}
\mathbf{h}_{0}    & \mathbf{0}      & \ldots     & \mathbf{0}      &   \mathbf{h}_{\nu} & \ldots           & \mathbf{h}_{2}  & \mathbf{h}_{1}    \\
\mathbf{h}_{1}    & \mathbf{h}_{0}  & \ldots     & \mathbf{0}      &  \mathbf{0}        & \mathbf{h}_{\nu} & \ldots          & \mathbf{h}_{2}    \\
\vdots            & \vdots          &            &                 &  \ldots            & \ldots           & \mathbf{0}      & \mathbf{0}        \\
\mathbf{0}        & \ldots          & \mathbf{0} & \mathbf{h}_{\nu}& \mathbf{h}_{\nu-1} & \ldots           &  \mathbf{h}_{1} & \mathbf{h}_{0}    
\end{array}
\right]
}_{\mathbf{H}}
\underbrace{
\left[ \begin{array}{c}
x[0] \\ x[1] \\ \vdots \\ x[N-1] \\ \mathbf{0}_{\nu \times 1}
\end{array}
\right]
}_{\left[\begin{array}{c} \mathbf{x} \\ 0 \end{array}\right]}
+
\underbrace{
\left[ \begin{array}{c}
\mathbf{z}[0] \\ \mathbf{z}[1] \\ \vdots \\ \mathbf{z}[N]  \\ \vdots \\ \mathbf{z}[N+\nu]
\end{array}
\right]
}_\mathbf{z}
\end{align}
\setcounter{equation}{\value{mytempeqncnt}}
\hrulefill
\vspace*{4pt}
\end{figure*}

For decoding the higher layer we treat the signal on the lower layer
as noise. Given that $\mathbf{h} \in \mathcal{A}^c_H$ and choosing
$\beta>\tilde{r}_H$ we conclude from lemma \ref{lemma3} that the
second term in (\ref{errorprob2}) decays exponentially in
$SNR$. Therefore,
\begin{align}
P_{e}^H(SNR) & \stackrel{\cdot}{\leq} \frac{1}{SNR^{(\nu+1)(1-\tilde{r}_H)}}
\end{align}
Or equivalently,
\begin{align}
(\nu+1)\left(1-\frac{N+\nu}{N}r_H\right) & \leq  d_{H}(r_H)
\end{align}

Once we have decoded the upper layer we subtract its contribution from
the lower layer. Proceeding as above, define
\begin{align}
\mathcal{A}_L & =  \left\{ \mathbf{h} :  |h_m|^2 \stackrel{\cdot}{\leq}\frac{1}{SNR^{1-\tilde{r}_L-\beta}} \quad \forall m \in \{0,1,\ldots,\nu\}  \right\} 
\end{align}
For high $SNR$ it follows that,
\begin{align*} 
P(\mathcal{A}_L)& =SNR^{-(\nu+1)(1-\tilde{r}_L-\beta)},\\ 
P(\mathcal{A}^c_L)& =1-SNR^{-(\nu+1)(1-\tilde{r}_L-\beta)}.
\end{align*}  
For the lower layer we have that
$(d_{min}^L)^2=\frac{SNR^{1-\beta}}{SNR^{\tilde{r}_L}}=SNR^{1-\beta-\tilde{r}_L}$. Using
lemma \ref{lemma1}, taking $\beta$ arbitrarily close to $\tilde{r}_H$
we can conclude that,
\begin{align}
(\nu+1)\left(1-\frac{N+\nu}{N}(r_{H}+r_L)\right) & \leq  d_{L}(r_{L}) \nonumber \\
&\hspace{-0.4in}  \leq  (\nu+1)\left(1-(r_H+r_L)\right) 
\end{align}
Comparing this with Theorem \ref{theorem:dtse} we can see that the
diversity multiplexing tradeoff for the ISI channel is successively
refinable since $d_H(r_H)=d_{isi}(r_H)$ and
$d_L(r_L)=d_{isi}(r_H+r_L)$.
\end{proof}

The intuition that was used in deriving the successive refinement of
the SISO tradeoff for ISI channels was that given that
$h\in\mathcal{A}$ at most $\nu$ taps in the frequency domain are zero
and the remaining are ``good'' and of the same magnitude. This
intuition can also be carried over to show the successive refinability
of the SIMO channel with $M_r$ receive antennas and one transmit
antenna. In this case, the received vector at the $n^{th}$ instant is
given by,
\begin{equation}
\label{eqn:sysmodel_simo}
\mathbf{y}[n] = \mathbf{h}_{0} x[n] + \mathbf{h}_{1} x[n-1] +\ldots +\mathbf{h}_{\nu}x[n-\nu] + \mathbf{z}[n]
\end{equation}
where $\mathbf{y},\mathbf{h}_{i},\mathbf{z} \in \mathbb{C}^{M_r
\times 1 }$. Assume that the $\nu+1$ fading coefficients are
$\mathbf{h}_{i} \sim
\mathcal{C}\mathcal{N}(\mathbf{0},\mathbf{I}_{M_r})$ and fixed for the
duration of the block length ($N+\nu$) and $\mathbf{h}_{i}$ is
independent of $\mathbf{h}_{j}$. Let $h_{i}^{(p)}$ represent the
$i^{th}$ tap coefficient between the transmitter and the $p^{th}$
receive antenna. We will denote
$\mathbf{C}=circ\{c_1,c_2,\ldots,c_T\}$ to be the $T \times T$
circulant matrix given by
\begin{align}
\label{def_circ}
\mathbf{C} & =  \left[ 
\begin{array}{cccccc}
c_1 & c_2 & c_3 & \ldots & c_{T-1} & c_T \\
c_T & c_1 & c_2 & \ldots & c_{T-2} & c_{T-1} \\
\vdots & & \vdots & \ddots &  &\vdots \\
c_2 & c_3 & c_4 & \ldots &  c_T  & c_1 \\
\end{array}
\right]
\end{align}

Consider a transmission scheme in which we transmit uncoded symbols
from a QAM constellation of size $SNR^r$ for $N$ time instants and pad
them with $\nu$ zero symbols at the end.  Consider a transmission
scheme in which one data symbol is sent at every instant from a QAM
constellation of size $SNR^r$. Therefore, the received vector over the
block of length $N+\nu$ can be written as in equation
(\ref{circulant_model_simo}) at the top of the page, where
$\mathbf{H}\in\mathbb{C}^{(N+\nu)M_r \times (N+\nu)}$,
$\mathbf{y}\in\mathbb{C}^{(N+\nu)M_r \times 1}$, $\mathbf{z}
\in\mathbb{C}^{(N+\nu)M_r \times 1}$ and $\mathbf{x}\in\mathbb{C}^{N
\times 1}$.  \addtocounter{equation}{1} By reordering the rows we can
write the received vector in terms of circulant matrices as,
\begin{align}
\label{simo_eqn1}
\left[ \begin{array}{c} \mathbf{y}^{(1)} \\ \mathbf{y}^{(2)} \\\vdots \\ \mathbf{y}^{(M_r)} \end{array} \right] & = 
\left[
\begin{array}{c}\mathbf{H}^{(1)} \\ \mathbf{H}^{(2)} \\ \vdots \\ \mathbf{H}^{(M_r)} \end{array}
\right] 
 \left[\begin{array}{l} \mathbf{x} \\ \mathbf{0}_{\nu \times 1} \end{array}\right]+
\left[ \begin{array}{c} \mathbf{z}^{(1)} \\ \mathbf{z}^{(2)} \\ \vdots \\ \mathbf{z}^{(M_r)} \end{array} \right]
\end{align}
where $\mathbf{H}^{(1)} , \ldots ,\mathbf{H}^{(M_r)} \in \mathbb{C}^{(N+\nu) \times (N+\nu) }$ are circulant
matrices given by,
\begin{align*}
\mathbf{H}^{(p)} & =  circ\{h_0^{(p)},0,\ldots,0,h_{\nu}^{(p)} ,\ldots,h_2^{(p)},h_1^{(p)}  \}
\end{align*}
for $p\in \{1,2,\ldots,M_r \}$ and,
\begin{align*}
\mathbf{y}^{(p)} & =  \left[ \begin{array}{c} y^{(p)}[0] \\ y^{(p)}[1] \\ \vdots \\ y^{(p)}[N+\nu]  \end{array} \right]
\end{align*}
where $y^{(p)}[n]$ represents the symbol received at the $p^{th}$
receive antenna in the $n^{th}$ time instant.

Since the $\mathbf{H}^{(p)}$ are circulant matrices we can write
them using the frequency domain notation as
$\mathbf{H}^{(p)}=\mathbf{Q}\mathbf{\Lambda}^{(p)} \mathbf{Q^*}$ where
$\mathbf{Q},\mathbf{Q^*}\in\mathbb{C}^{(N+\nu)\times(N+\nu)}$ are
truncated DFT matrices as defined earlier and $\mathbf{\Lambda}^{(p)}$
are diagonal matrices with the elements given by,
\begin{align*}
\mathbf{\Lambda}^{(p)} & =  diag\left\{\Lambda_{k}^{(p)} : \Lambda_{k}^{(p)}=\sum_{m=0}^{\nu}h_m^{(p)} e^{-\frac{2 \pi j}{(N+\nu)}k m}  \right\}
\end{align*}
for $ k =\{0,\ldots,(N+\nu-1)\}$. Therefore the equation (\ref{simo_eqn1}) can be rewritten as,
\begin{align}
\left[ \begin{array}{c} \mathbf{y}^{(1)} \\ \mathbf{y}^{(2)} \\\vdots \\ \mathbf{y}^{(M_r)} \end{array} \right] & = 
\left[
\begin{array}{c}\mathbf{H}^{(1)} \\ \mathbf{H}^{(2)} \\  \vdots \\ \mathbf{H}^{(M_r)} \end{array}
\right]  
\left[ \begin{array}{l}  \mathbf{x} \\ \mathbf{0}_{\nu \times 1}  \end{array}\right]
+ \left[ \begin{array}{c} \mathbf{z}^{(1)} \\ \mathbf{z}^{(2)} \\ \vdots \\ \mathbf{z}^{(M_r)} \end{array} \right]\\
& = 
\left[
\begin{array}{c}
\mathbf{Q}\mathbf{\Lambda}^{(1)} \mathbf{Q^*}  \\ \mathbf{Q}\mathbf{\Lambda}^{(2)} \mathbf{Q^*}  \\ \vdots \\ \mathbf{Q}\mathbf{\Lambda}^{(M_r)} \mathbf{Q^*} 
\end{array}
\right]    \left[\begin{array}{l} \mathbf{x} \\ \mathbf{0}_{\nu \times 1} \end{array}\right] + \left[ \begin{array}{c} \mathbf{z}^{(1)} \\ \mathbf{z}^{(2)} \\ \vdots \\ \mathbf{z}^{(M_r)} \end{array} \right]
\\
& = 
%\left[\begin{array}{cccc} 
%\mathbf{Q} & 0 & \ldots & 0 \\
%0 & \mathbf{Q} & 0 & \ldots \\
%\vdots &  & \ddots & \vdots \\
%0 & \ldots & 0 & \mathbf{Q} 
%\end{array}\right]
%\left[\begin{array}{c} 
%\mathbf{\Lambda}^{(1)} \\
%\mathbf{\Lambda}^{(2)} \\
%\vdots \\
%\mathbf{\Lambda}^{(M_r)} 
%\end{array}\right]
%\mathbf{\tilde{Q}^*} \mathbf{x} 
\left[
\begin{array}{c}
\mathbf{Q}\mathbf{\Lambda}^{(1)} \mathbf{\tilde{Q}^*} \mathbf{x}  \\ \mathbf{Q}\mathbf{\Lambda}^{(2)} \mathbf{\tilde{Q}^*} \mathbf{x} \\ \vdots \\ \mathbf{Q}\mathbf{\Lambda}^{(M_r)} \mathbf{\tilde{Q}^*} \mathbf{x}
\end{array}
\right]   + \left[ \begin{array}{c} \mathbf{z}^{(1)} \\ \mathbf{z}^{(2)} \\ \vdots \\ \mathbf{z}^{(M_r)} \end{array} \right]
\end{align}
where $\mathbf{\tilde{Q}^*}$ is a $(N+\nu)\times N $ matrix which is
obtained by deleting the last $\nu$ columns of the matrix
$\mathbf{Q}^*$ (similar as in the proof of lemma \ref{lemma1}).

\begin{lemma}
\label{magnitude_simo}
For a $(\nu+1)$ tap, $M_r$ receive antennas ISI channel we have the
taps in the frequency domain are given by,
\begin{align*}
\Lambda_{k}^{(p)} & =  \sum_{m=0}^{\nu}h_m^{(p)} e^{-\frac{2 \pi j}{(N+\nu)}k m}  
\end{align*}
for $k =\{0,\ldots,(N+\nu-1)\}$ and $\ p \in\{0,\ldots,M_r\}$. Define
the sets $\mathcal{F}^{(p)}$, $\mathcal{G}^{(p)}$ and $\mathcal{M}$
as,
\begin{align} 
\mathcal{F}^{(p)} & =  \{k : |\Lambda_k^{(p)}|^2\stackrel{\cdot}{\leq} SNR^{-(1-r)}\},\\
\mathcal{G}^{(p)} & =  \{k : |\Lambda_k^{(p)}|^2 \doteq \max_{l\in\{ 0,1,\ldots,\nu\} }|h_l^{(p)}|^2 \} \\
\mathcal{M} & =  \{ \mathbf{h} :  |h_{i}^{(p)}|^2 \stackrel{\cdot}{\leq}\frac{1}{SNR^{1-r}} \nonumber \\
& \hspace{0.2in}  \forall i \in \{0,\ldots,\nu\} , \ \forall p \in \{1,\ldots,M_r\}  \}  
\end{align}
With $\overline{\mathcal{G}^{(p)}}$ representing the complement of the
set $\mathcal{G}^{(p)}$, we have
\begin{align*}
  |\overline{\mathcal{G}^{(p)}}|& \leq \nu \quad \forall p
\end{align*}
and given that $\mathbf{h} \in \mathcal{M}^c$ this means that,
\begin{equation}
\exists p \in \{1,2,\ldots,M_r\} \ s.\ t.\   |\mathcal{F}^{(p)}|\leq \nu
\end{equation}
\qed
\end{lemma}
\begin{proof}
From lemma \ref{magnitude} for each $p$ it is clear that
$|\overline{\mathcal{G}^{(p)}}| \leq \nu$. Since $\mathbf{h} \in \mathcal{M}^c$
there exists at least one $(i,p)$ pair such that $|h_{i}^{(p)}|^2
\stackrel{\cdot}{\leq}\frac{1}{SNR^{1-r}}$. Then from lemma
\ref{magnitude} it follows that for this particular $p$,
$|\mathcal{F}^{(p)}|\leq \nu$ and $|\overline{\mathcal{G}^{(p)}}|\leq \nu$.
\end{proof}

As before, define
\begin{align}
\label{M_high}
\mathcal{M}_H & =   \{ \mathbf{h} :  |h_{i}^{(p)}|^2 \stackrel{\cdot}{\leq}\frac{1}{SNR^{1-r_H}} \nonumber \\
&\hspace{0.2in}  \forall i \in \{0,1,\ldots,\nu\} , \  \forall p \in \{1,2,\ldots,M_r\}  \}  
\end{align}
Since all the tap coefficients are i.i.d we have that
$P(\mathcal{M})=SNR^{-M_r(\nu+1)(1-r)}$ and
$P(\mathcal{M}_H)=SNR^{-M_r(\nu+1)(1-r_H)}$.

\begin{lemma}
\label{lemma3_simo}
Using the $\mathcal{X}_H$ and $\mathcal{X}_L$ defined earlier for
signaling, assume uncoded superposition transmission such that at each
time instant one symbol is independently transmitted from each
constellation ($\mathcal{X}_H$, $\mathcal{X}_L$) for $N$ time instants
followed by a padding with $\nu$ zero symbols. For a finite period of
communication (finite $N$) given that $\mathbf{h} \in \mathcal{M}_H^c$
(see (\ref{M_high})), the error probability of detecting the set of
symbols sent from the higher constellation ($\mathcal{X}_H$) denoted
by $P_e^{H}(SNR)$ decays exponentially in $SNR$.  \qed
\end{lemma}
We will just give an outline of the proof as the details are
similar to the proof of lemma \ref{lemma3} given in the Appendix. From
lemma \ref{magnitude_simo} there exists at least one set of
$(N+\nu)$ coefficients in the frequency domain through which
$\mathbf{\hat{Q}}\mathbf{x}$ passes such that at most $\nu$ taps of
the available $(N+\nu)$ taps in this set are of magnitude smaller than
$SNR^{-(1-r_H)}$. Then from lemma \ref{lemma3} it directly follows
that the error probability decays exponentially in $SNR$.

\begin{theorem}
Consider a $\nu$ tap point to point SIMO ISI channel with $M_r$
receive antennas. The diversity multiplexing tradeoff for this channel
is successively refinable, {\em i.e.}, for any multiplexing gains
$r_H$ and $r_L$ such that $r_H+r_L \leq \frac{N}{N+\nu}$ the
achievable diversity orders given by $d_{H}(r_H)$ and
$d_{L}(r_L)$ are bounded as,
\begin{align}
M_r (\nu+1)\left(1-\frac{N+\nu}{N}r_H\right) & \leq  d_{H}(r_H) \nonumber \\
& \hspace{-0.6in} \leq  M_r (\nu+1)\left(1-{r_H}\right) , \label{higherlayer_simo}\\
M_r (\nu+1)\left(1-\frac{N+\nu}{N}(r_{H}+r_L)\right) & \leq  d_{L}(r_{L}) \nonumber \\
& \hspace{-0.6in} \leq  M_r (\nu+1)\left(1-(r_H+r_L)\right)  \label{higherlower_simo}
\end{align}
where $N$ is finite and does not grow with SNR. 
\qed
\end{theorem}
\begin{proof} As in theorem \ref{theorem_siso} use superposition coding and assume two streams with uncoded QAM codebooks $\mathcal{X}_H$ and
$\mathcal{X}_L$ for the higher and lower priority streams
respectively.  Choose $SNR_{H}\doteq SNR$ and $SNR_{L} \doteq
SNR^{1-\beta}$ for $\beta \in [0,1]$. Also let
$|\mathcal{X}_H|=SNR^{\tilde{r}_H}$ and
$|\mathcal{X}_L=SNR^{\tilde{r}_L}$. As in theorem \ref{theorem_siso}
the minimum distances are $(d_{min}^H)^2=SNR^{1-\tilde{r}_H}$,
$(d_{min}^L)^2=SNR^{1-\beta-\tilde{r}_L}$, and if $\beta>\tilde{r}_H$.
the order of magnitude of the effective minimum distance between two
points in the constellation $\mathcal{X}_H$ is preserved.

The symbol transmitted at the $k^{th}$ instant is
the superposition of a symbol from $\mathcal{X}_H$, $\mathcal{X}_L$
given by,
\begin{align*}
x[k] & =  x_H[k]+x_L[k] \qquad {\rm where} \, \, x_H[k] \in \mathcal{X}_H,  \,\, x_L[k] \in \mathcal{X}_L
\end{align*}

The upper bound in both (\ref{higherlayer_simo}) and (\ref{higherlower_simo}) is
trivial and follows from the matched filter bound. We will investigate
the lower bound in (\ref{higherlayer_simo}). At each time instant superpose
symbols from the higher and lower layers for $N$ time instants and pad
them with $\nu$ zero symbols at the end.  We consider this particular
transmission scheme and for detection, given that $\mathbf{h} \in
\mathcal{M}^c_H $, (where $\mathcal{M}_H$ is as defined in equation
(\ref{M_high})), we proceed as in lemma 3. Therefore, we can write,
\begin{align}
\lefteqn{P_{e}^H(SNR)  =   P(\mathcal{M}_H) P_{e}(SNR \mid \mathbf{h} \in \mathcal{M}_H )}&\nonumber \\
&\hspace{0.2in} + P(\mathcal{M}^c_H) P_{e}(SNR\mid \mathbf{h} \in \mathcal{M}^c_H ) \nonumber \\
%& \stackrel{\cdot}{\leq}  P(\mathcal{M}_H) + \left( 1-SNR^{-M_r (\nu+1)(1-\tilde{r}_H)} \right)  P_{e}(SNR\mid \mathbf{h} \in \mathcal{M}^c_H )  \nonumber \\
& \stackrel{\cdot}{\leq}  SNR^{-M_r (\nu+1)(1-\tilde{r}_H)} + \\
&\hspace{0.2in} \left( 1-SNR^{-M_r(\nu+1)(1-\tilde{r}_H)} \right)   P_{e}^{D}(SNR\mid \mathbf{h} \in \mathcal{M}^c_H ). \label{errorprob2_simo}
\end{align}
for communication at an effective rate of $r_H=\frac{N}{N+\nu}\tilde{r}_H$.

From lemma \ref{lemma3_simo}, where we treat the signal on the lower
layer as noise, we get that the second term in (\ref{errorprob2_simo})
decays exponentially in $SNR$. Therefore,
\begin{align}
P_{e}^H(SNR) & \stackrel{\cdot}{\leq}   \frac{1}{SNR^{M_r (\nu+1)(1-\tilde{r}_H)}}
\end{align}
Or equivalently,
\begin{align}
M_r (\nu+1)\left(1-\frac{N+\nu}{N}r_H\right) & \leq  d_{H}(r_H)
\end{align}

Once we have decoded the upper layer we subtract its contribution from
the lower layer. Proceeding as above, define
\begin{align}
\mathcal{M}_L & =  \{ \mathbf{h} :  |h_i^{(p)}|^2 \stackrel{\cdot}{\leq}\frac{1}{SNR^{1-\tilde{r}_L-\beta}} \nonumber \\
& \hspace{0.2in} \forall i \in \{0,1,\ldots,\nu\} , \ \forall p \in \{1,2,\ldots,M_r\}  \} 
\end{align}
For high $SNR$ it follows that $P(\mathcal{M}_L)=SNR^{-M_r
(\nu+1)(1-\tilde{r}_L-\beta)}$. Using lemma \ref{lemma3_simo}, taking
$\beta$ arbitrarily close to $\tilde{r}_H$ we can conclude that,
\begin{align}
M_r (\nu+1)\left(1-\frac{N+\nu}{N}(r_{H}+r_L)\right) & \leq   d_{L}(r_{L}) \nonumber \\
&\hspace{-0.6in} \leq M_r (\nu+1)\left(1-(r_H+r_L)\right) 
\end{align}
Comparing this with Theorem \ref{theorem:dtse} we can see that the
diversity multiplexing tradeoff for the ISI channel is successively
refinable.
\end{proof}

\section{Discussion}
In this paper we presented the successive refinement of the diversity
multiplexing tradeoff for the SISO ISI fading channel. Moreover we
showed that superposition of two uncoded QAM constellations was
sufficient to achieve this successive refinement.  Although parallel
channels are not successively refinable, a set of correlated parallel
channels might be refinable.  The same result holds for multiple
receive and single transmit antenna. It would be interesting to
investigate whether a similar result would be true for ISI channels
with a single receive and multiple transmit antennas.

\section{Appendix}

\subsection{Proof of lemma \ref{magnitude}}
\begin{proof}
The tap coefficients in the frequency domain are given by,
\begin{align}
\Lambda_{k} & =    \sum_{m=0}^{\nu} h_m e^{-\frac{2 \pi j}{(N+\nu)}k m} \qquad k =\{0,\ldots,(N+\nu-1)\}
\nonumber
\end{align}
Defining $\theta=e^{-\frac{2 \pi j}{(N+\nu)}}$ the above equation can be rewritten as,
\begin{align}
\Lambda_{k} & =    \sum_{m=0}^{\nu} h_m \theta ^{k m} \qquad k =\{0,\ldots,(N+\nu-1)\} \\ \nonumber
& =  \left[\begin{array}{cccc} 1 & \theta^{k} & \ldots & \theta^{k\nu} \end{array} \right] \left[\begin{array}{cccc} h_0 & h_1 & \ldots & h_\nu \end{array}\right]^t
\nonumber
\end{align}

Take any set of $(\nu+1)$ coefficients in the frequency domain and
index this set by $\mathcal{K}=\{k_0,\ldots,k_{\nu} \}$. Define,
\begin{align}
\label{eqn:use_inverse}
\mathbf{\Breve{\Lambda}} & =  \left[\begin{array}{c} \Lambda_{k_0} \\\Lambda_{k_1} \\ \ldots \\ \Lambda_{k_{\nu}} \end{array}\right] 
= \underbrace{\left[ 
\begin{array}{cccc} 
1 & \theta^{k_0} & \ldots & \theta^{k_0\nu} \\
1 & \theta^{k_1} & \ldots & \theta^{k_1\nu} \\
\vdots & & \ldots & \vdots \\
1 & \theta^{k_{\nu}} & \ldots & \theta^{k_{\nu}\nu} \\
\end{array}
\right] }_{\mathbf{V}} 
\underbrace{
\left[\begin{array}{c} h_0 \\ h_1 \\ \ldots \\ h_\nu \end{array}\right]
}_{\mathbf{h}} 
\end{align}
where $\mathbf{V}\in\mathbb{C}^{(\nu+1)\times(\nu+1)}$ is a full rank
Vandermonde matrix. Therefore its inverse exists and we denote it by
$\mathbf{V}^{-1}=\mathbf{A}$. Denoting the rows of $\mathbf{A}$ as,
\begin{align}
\mathbf{A} & =  \left[ \begin{array}{c} \mathbf{a}^{(0)} \\ \mathbf{a}^{(1)}
 \\ \ldots  \\ \mathbf{a}^{(\nu)} \end{array} \right]
\end{align}
 we conclude that,
\begin{align}
\label{eqn:unit_vector}
\mathbf{a}^{(l)} \mathbf{V}  & =  \mathbf{e}^{(l)} \quad \mbox{ and }\  \mathbf{a}^{(l)} \neq \mathbf{0}
\end{align}
where $\mathbf{e}^{(l)} \in \mathbb{C}^{1\times (\nu+1)}$ is the unit
row vector with $1$ at the $l^{th}$ position and zero otherwise. Note
that the entries of $\{ \mathbf{a}^{(l)}\}$ do not depend on
$SNR$. Therefore,
\begin{align}
\label{eqn:unit_vector_use}
\mathbf{a}^{(l)} & =  \mathbf{e}^{(l)} \mathbf{V}^{-1}
\end{align}
From (\ref{eqn:use_inverse}) we have,
\begin{align*}
\mathbf{h} &=  \mathbf{V}^{-1} \Breve{\mathbf{\Lambda}} 
\end{align*}
Multiplying both sides by $\mathbf{e}^{(l)}$ and using
(\ref{eqn:unit_vector_use}) we get,
\begin{align*}
\mathbf{e}^{(l)} \mathbf{h} & =  h_l = \mathbf{e}^{(l)} \mathbf{V}^{-1} \Breve{\mathbf{\Lambda}} \stackrel{(a)}{=} \mathbf{a}^{(l)} \Breve{\mathbf{\Lambda}} =  \sum_{i=0}^{\nu} a_{i}^{(l)} \Lambda_{k_i} 
\end{align*}
Using the Cauchy-Schwartz inequality\footnote{$|\mathbf{u}^* \mathbf{v}|\leq \|u \|. \|v\|$.}, we get,
\begin{align*}
|h_l|^2 & = |\sum_{i=0}^{\nu} a_{i}^{(l)} \Lambda_{k_i}|^2 \leq (\sum_{i=0}^{\nu} |a_{i}^{(l)} |^2)  (\sum_{i=0}^{\nu} |\Lambda_{k_i} |^2) 
\end{align*}
Using the fact that $N$ is finite or does not grow with $SNR$ it
follows that the $\{a_{i}^{(l)}\}$ do not depend on
$SNR$. Therefore, the above inequality can be written as
\begin{align}
\label{cauchy}
|h_l|^2  & \stackrel{\cdot}{\leq}   |\Lambda_{k_0}|^2 + |\Lambda_{k_1}|^2 + \ldots + |\Lambda_{k_\nu}|^2  
\end{align}
Note that the above inequality holds for all $h_l$,
$l=0,\ldots,\nu$. Therefore, we get that for any set of $(\nu+1)$
coefficients in the frequency domain indexed by $\{k_0,\ldots,k_{\nu}
\}$,
\begin{align}
\label{cauchy1}
\max_{l\in\{ 0,1,\ldots,\nu\} }|h_l|^2  & \stackrel{\cdot}{\leq}   |\Lambda_{k_0}|^2 + |\Lambda_{k_1}|^2 + \ldots + |\Lambda_{k_\nu}|^2  
\end{align}
From the Cauchy-Schwartz inequality note that,
\begin{align}
\lefteqn{|\Lambda_{k_0}|^2 + |\Lambda_{k_1}|^2 + \ldots + |\Lambda_{k_\nu}|^2   = |\sum_{m=0}^{\nu} h_m \theta ^{k_0 m}|^2 +\ldots}&\nonumber\\
 & \hspace{1in} + |\sum_{m=0}^{\nu} h_m \theta ^{k_\nu m}|^2  \nonumber \\
\leq &  (\sum_{m=0}^{\nu} |h_m|^2 )\left(\sum_{m=0}^{\nu}|\theta ^{k_0 m}|^2 + \ldots +  \sum_{m=0}^{\nu}|\theta ^{k_{\nu} m}|^2\right)\nonumber \\
%= & (\nu+1)^2 (|h_0|^2+|h_1|^2+\ldots+|h_{\nu}|^2 )  \nonumber \\
\doteq &  (|h_0|^2+|h_1|^2+\ldots+|h_{\nu}|^2 ) \nonumber \\
\doteq &  \max_{l\in\{ 0,1,\ldots,\nu\} }|h_l|^2 \label{cauchy2}
\end{align}
Combining equations (\ref{cauchy1}) and (\ref{cauchy2}) we get,
\begin{align}
\label{cauchy3}
 |\Lambda_{k_0}|^2 + |\Lambda_{k_1}|^2 + \ldots + |\Lambda_{k_\nu}|^2 & \doteq  \max_{l\in\{ 0,1,\ldots,\nu\} }|h_l|^2 .
\end{align}
\begin{itemize}
\item Given that $\mathbf{h}\in \mathcal{A}^c$ we know that there exists at
least one $l$ such that $|h_l|^2 \stackrel{\cdot}{\geq}
SNR^{-(1-r)}$. Therefore, if more than $\nu$ taps in the frequency
domain are of magnitude less than $SNR^{-(1-r)}$ choose our set $\mathcal{K}$ to
be these sets of coefficients. From equation (\ref{cauchy3}) we get a
contradiction. Therefore $|\mathcal{F}|\leq \nu$ proving $(a)$.
\item We know from (\ref{cauchy2}) that,
\begin{align*}
 |\Lambda_k|^2 & \stackrel{\cdot}{\leq} \max_{l\in\{ 0,1,\ldots,\nu\} }|h_l|^2 \qquad  \forall k
\end{align*}
Since, $ \mathcal{G} =\{i : |\Lambda_i|^2 \doteq \max_{l\in\{
0,1,\ldots,\nu\} }|h_l|^2\}$,
\begin{align*}
 |\Lambda_k|^2 & \stackrel{\cdot}{<} \max_{l\in\{ 0,1,\ldots,\nu\}  }|h_l|^2 \qquad \forall k \in \mathcal{G}^c.
\end{align*}
If $|\mathcal{G}^c|>\nu $ then there exists a set
 $\mathcal{K}=\mathcal{G}^{c}$ of size at least $\nu+1$ such that,
\begin{align*}
 |\Lambda_k|^2 & \stackrel{\cdot}{<}  \max_{l\in\{ 0,1,\ldots,\nu\} }|h_l|^2 \qquad  \forall k \in \mathcal{K}
\end{align*}
But this is a contradiction to equation (\ref{cauchy3}) and
therefore we have $|\mathcal{G}^c|\leq \nu$ proving $(b)$.
\end{itemize}

\end{proof}

\subsection{Proof of lemma \ref{lemma1}}
\begin{proof}
Consider the case where we have $\nu+1$ taps, {\em i.e.},
\begin{align*}
y[n] & =   \sum_{m=0}^{\nu} h_m x[n-m]+ z[n] 
\end{align*}
We receive a vector of length $(N+\nu)$ denoted by
$\mathbf{y}$. Denoting the transmitted sequence of length $N$ by
$\mathbf{x}\in \mathcal{X}^{N}$, the $\nu$ zero symbols padded at the
end by $\mathbf{0}_{\nu \times 1}$ and the circulant channel matrix as
$\mathbf{H}$, we have
\begin{align}
\mathbf{y} & =  \mathbf{H} \left[\begin{array}{cc} \mathbf{x} & \mathbf{0}_{\nu \times 1} \end{array}\right]^t+\mathbf{z} \label{use_for_simo}
\end{align}
Similar to analysis \cite{dtse} we can write the circulant matrix
$\mathbf{H=Q}\mathbf{\Lambda}\mathbf{ Q^*}$ where ${\bf Q} \in
\mathbb{C}^{(N+\nu)\times(N+\nu)} $ and ${\bf Q^*} \in
\mathbb{C}^{(N+\nu)\times(N+\nu)} $ are truncated DFT matrices and
$\mathbf{\Lambda}\in \mathbb{C}^{(N+\nu)\times(N+\nu)}$ is a diagonal matrix with the diagonal elements
given by,
\begin{align*}
\Lambda_{k} & =   \sum_{m=0}^{\nu} h_m e^{-\frac{2 \pi j}{(N+\nu)} k m} \qquad k =\{0,\ldots,(N+\nu-1)\}
\end{align*}
and the entries of $\mathbf{Q}$ are given by,
\begin{align*}
(\mathbf{Q})_{pq} & =  e^{-\frac{2 \pi j}{(N+\nu)}pq} \ {\rm for} \,\, 0 \leq p \leq (N+\nu), \,\, 0\leq q \leq (N+\nu)
\end{align*}
Note that $\mathbf{Q}$ is a Vandermonde matrix. Multiplying the
received vector by $\mathbf{Q^*}$ we get,
\begin{align*}
\mathbf{\tilde{y}} & =  \mathbf{Q^* y} = \mathbf{\Lambda} \mathbf{Q^*\left[\begin{array}{cc} \mathbf{x} & \mathbf{0}_{\nu \times 1} \end{array}\right]^t } + \mathbf{ Q z} = \mathbf{\Lambda} \mathbf{\tilde{Q}^*} \mathbf{x} + \mathbf{\tilde{z}} 
\end{align*}
where $\mathbf{\tilde{Q}^*}\in \mathbb{C}^{(N+\nu)\times N} $ is a matrix
obtained by deleting the last $\nu$ columns. Note that
$\mathbf{\tilde{Q}^*}$ is also a Vandermonde matrix which implies it
has rank $N$.

Note that from lemma \ref{magnitude} we know that at most $\nu$ taps
of the available $(N+\nu)$ taps in the frequency domain can be of
magnitude $|\Lambda_{k}|^2\stackrel{\cdot}{\leq} SNR^{-(1-r)}$. Define
a selection matrix $\mathbf{S} \in \mathbb{C}^{N \times (N+\nu)}$ such that,
\begin{align*}
  \mathbf{S}\mathbf{\Lambda} \mathbf{\tilde{Q}}^* &  =  \bfhat{\Lambda} \bfhat{Q}
\end{align*}
where, $\bfhat{\Lambda}\in\mathbb{C}^{N \times N}$,
$\bfhat{\Lambda}=diag\left( \{\Lambda_l:l\in
\mathcal{F}^c\}\right)$. Similarly $\bfhat{Q}\in \mathbb{C}^{N \times
N}$ is the matrix $\mathbf{\tilde{Q}}$ with the $\nu$ rows
corresponding to $\{\Lambda_l:l\in \mathcal{F}\}$ deleted. Note that
$\bfhat{Q}$ is still a full rank (rank $N$) Vandermonde matrix and
denoting the singular values of $\bfhat{\Lambda} \bfhat{Q}$ by
$\gamma_k$ we have $\gamma_k \stackrel{\cdot}{>}SNR^{-(1-r)} $. Using
this selection matrix we have,
\begin{align}
\label{lemma2_eqn1}
\bfhat{y} & =  \mathbf{S} \mathbf{\tilde{y}} = \bfhat{\Lambda} \bfhat{Q}\mathbf{x} + \bfhat{z}
\end{align}
Due to the fact that we are using uncoded QAM for transmission, the
minimum norm distance between any two elements
$\mathbf{x}\neq\mathbf{x'} \in \mathcal{X}^{N}$ is lower bounded by,
\begin{align*}
\|\mathbf{x}-\mathbf{x'}\|^2 & \stackrel{\cdot}{\geq}  SNR^{(1-r)}.
\end{align*}
From the fact that $\bfhat{Q}$ is full rank its smallest
singular value is nonzero and independent of SNR. Defining
$\bfhat{x}=\bfhat{Q} \mathbf{x}$ we can conclude that,
\begin{align}
\label{distance_preserved}
\|\bfhat{x} - \bfhat{x}'\|^2 & \doteq   \|\mathbf{x}-\mathbf{x'}\|^2  \stackrel{\cdot}{\geq} SNR^{(1-r)}
\end{align}
As $\bfhat{\Lambda}$ is a diagonal matrix, 
\begin{align}
\|\bfhat{\Lambda} (\bfhat{x}-\bfhat{x}') \|^2 & = \sum_{l=0}^{N-1} | \Lambda_l (\bfhat{x} - \bfhat{x}')_l |^2 =   \sum_{l=0}^{N-1} | \Lambda_l |^2  | (\bfhat{x} - \bfhat{x}')_l |^2 \nonumber \\
& \doteq  SNR^{-(1-r)+\epsilon} \sum_{l=0}^{N-1} | (\bfhat{x} - \bfhat{x}')_l |^2 \label{use_lemma}\\
& =   SNR^{-(1-r)+\epsilon} \| (\bfhat{x}-\bfhat{x}') \|^2 \nonumber \\
& \stackrel{\cdot}{\geq}   SNR^{-(1-r)+\epsilon} SNR^{(1-r)} = SNR^{\epsilon} \nonumber
\end{align}
where (\ref{use_lemma}) is true from lemma \ref{magnitude} for some
$\epsilon>0$. Since $Q(x)$ is a decreasing function in $x$, using the
above equation, we conclude that the pairwise error probability of
detecting the sequence $\mathbf{x'}$ given that $\mathbf{x}$ was
transmitted is upper bounded by,
\begin{align}
P_e(\mathbf{x}\rightarrow \mathbf{x'}) & \leq  Q\left( \|\bfhat{\Lambda} (\bfhat{x}-\bfhat{x}') \|^2\right)  \stackrel{\cdot}{\leq} Q  \left( SNR^{\epsilon}   \right) \nonumber
\end{align}
Therefore, by the union bound we have,
\begin{align*}
P_e(SNR) & \stackrel{\cdot}{\leq}  SNR^r Q  \left( SNR^{\epsilon}  \right) \stackrel{\cdot}{\leq} SNR^r e^{-\frac{SNR^{2\epsilon}}{2}}
\end{align*}
as $Q(x)$ decays exponentially in $x$ for large $x$ {\em i.e.}, $Q(x)
 \leq e^{-\frac{x^2}{2}}$. Hence it follows that given $\mathbf{h} \in
 \mathcal{A}^c$ the error probability decays exponentially in
 $SNR$. Note that we only use the weaker form of lemma \ref{magnitude}
 over here {\em i.e.} we need at least $N$ tap coefficients to be
 large but we don't need them to be of the same magnitude.
\end{proof}

\subsection{Proof of lemma \ref{lemma3}}
\begin{proof}
For decoding the higher layer we treat the signal on the lower layer
as noise.  Proceed as in the previous lemma (equation
\ref{lemma2_eqn1}) with the selection matrix $\mathbf{S}$ chosen such
that $\bfhat{\Lambda}=diag\left( \{\Lambda_l:l\in
\mathcal{G}\}\right)$, where $|\mathcal{G}|\geq N$. We get,
\begin{align*}
\bfhat{y} & =  \mathbf{S} \mathbf{\tilde{y}} = \bfhat{\Lambda} \underbrace{\bfhat{Q}\mathbf{x_H}}_{\bfhat{x}_H}+\bfhat{\Lambda}\underbrace{ \bfhat{Q}\mathbf{x_L}}_{\bfhat{x}_L} + \bfhat{z} \\
& =   \bfhat{\Lambda}\bfhat{x}_H + \underbrace{\bfhat{\Lambda}\bfhat{x}_H+\bfhat{z}}_\mathbf{\tilde{z}} \\
& =  \bfhat{\Lambda}\bfhat{x}_H+ \mathbf{\tilde{z}} 
\end{align*}
The decoding rule we use to decode $\mathbf{x}_H$ is given by,
\begin{align*}
\mathbf{\tilde{x}}_H & =  \argmin_{\mathbf{x}_H} \| \bfhat{y}-\bfhat{\Lambda}\bfhat{Q} \mathbf{x}_H \|^2
\end{align*}
Therefore, the pairwise error probability of detecting the sequence
$\mathbf{x_H^{'}}$ given that $\mathbf{x_H}$ was transmitted is given by,
\begin{align}
\lefteqn{P_e^H(\mathbf{x}_H \rightarrow \mathbf{x}_H^{'})  =  \sum_{\mathbf{x}_L \in \mathcal{X}_L^N } Pr(\mathbf{x}_L) P_e(\mathbf{x_H}\rightarrow \mathbf{x}_H^{'} | \Lambda,\mathbf{x}_L )}  & \nonumber \\
& =  \sum_{\mathbf{x}_L \in \mathcal{X}_L^N } Pr(\mathbf{x}_L) Pr\left( \|\bfhat{y}-\bfhat{\Lambda} \bfhat{x}_H \|^2 > \|\bfhat{y}-\bfhat{\Lambda} \bfhat{x}_H^{'} \|^2 \right) \nonumber \\
& =  \sum_{ \mathbf{x}_L \in \mathcal{X}_L^N  } Pr(\mathbf{x}_L) Q \left( \| \bfhat{\Lambda}(\bfhat{x}_H-\bfhat{x}_H^{'})\| +\right.  \nonumber \\
& \hspace{1.4in} \left. 2 Re\frac{<\bfhat{\Lambda}(\bfhat{x}_H-\bfhat{x}_H^{'}),\bfhat{\Lambda}\bfhat{x}_L> }{\| \bfhat{\Lambda}(\bfhat{x}_H-\bfhat{x}_H^{'})\|} \right) \label{pep}
\end{align}
Note that $Q(x)$ is a decreasing function in $x$. Therefore the
equation (\ref{pep}) is upper bounded by,
\begin{align}
\lefteqn{P_e^H(\mathbf{x}_H \rightarrow \mathbf{x}_H^{'}) \leq} & \nonumber \\
&\hspace{0.4in}\sum_{ \mathbf{x}_L \in \mathcal{X}_L^N  } Pr(\mathbf{x}_L) Q \left(  \underbrace{\| \bfhat{\Lambda}(\bfhat{x}_H-\bfhat{x}_H^{'})\| - 2\|\bfhat{\Lambda}\bfhat{x}_L\|}_\Omega  \right) 
\end{align}
\newpage 
Define $\Gamma_{min}$ and $\Gamma_{max}$ as,
\begin{align*}
\Gamma_{min} & =  \min_{i\in \mathcal{G}} |\Lambda_i|^2, & \Gamma_{max} & =  \max_{i\in \mathcal{G}} |\Lambda_i|^2 
\end{align*}
Therefore, from lemma \ref{magnitude}, we get
\begin{align*}
\Gamma_{min} & \doteq  \Gamma_{max} \doteq  \max_{l\in\{ 0,1,\ldots,\nu\} }|h_l|^2 \doteq  SNR^{-(1-\tilde{r}_H)+ 2 \epsilon}
\end{align*}
where the last equality follows for some $\epsilon>0$ from lemma
\ref{magnitude} as $\mathbf{h}\in \mathcal{A}_H^c$. Since
$\|\bfhat{x}_L\|^2 \stackrel{\cdot}{\leq} SNR^{1-\beta}$ and from
equation (\ref{distance_preserved}) in the previous lemma, we
can lower bound $\Omega$ as,
\begin{align*}
\Omega & \geq  \Gamma_{min}^{\frac{1}{2}}  \| (\bfhat{x}_H-\bfhat{x}_H^{'})\| - 2 \Gamma_{max}^{\frac{1}{2}} \|\bfhat{x}_L\| \\
& \doteq  SNR^{\frac{-(1-\tilde{r}_H)+ 2 \epsilon}{2}}\left( \| \bfhat{x}_H-\bfhat{x}_H^{'}\| - \|\bfhat{x}_L\|\right) \\
& \doteq SNR^{-\frac{(1-\tilde{r}_H)}{2}+\epsilon}\left( SNR^{\frac{1-\tilde{r}_H}{2}}-SNR^{\frac{1-\beta}{2}}  \right) \\
& \doteq SNR^{\epsilon} 
\end{align*}
where the last step is valid as $\beta>\tilde{r}_H$
Therefore,
\begin{align*}
P_e^H (\mathbf{x}_H \rightarrow \mathbf{x}_H^{'}) & \stackrel{\cdot}{\leq} Q(SNR^{\epsilon} )
\end{align*}
which decays exponentially in SNR. By the union bound as in
the previous lemma we conclude that $P_e^{H}(SNR)$ decays
exponentially in SNR.

\end{proof}

\end{document}